\documentstyle[sprocl]{article}
\input epsf

\def\be{\begin{equation}}
\def\ee{\end{equation}}
\def\bea{\begin{eqnarray}}
\def\eea{\end{eqnarray}}

\begin{document}

\title{Lectures on Cosmic Topological Defects}

\author{Tanmay Vachaspati}
\address{
Department of Astronomy and Astrophysics,\\
T.I.F.R., Homi Bhabha Road,\\
Colaba, Mumbai 400 005, India\\
and\\
Physics Department,\\
Case Western Reserve University,\\
Cleveland, OH 44106-7079, USA.
}

\maketitle

\abstracts{
These lectures review certain topological defects and 
aspects of their cosmology. Unconventional material includes 
brief descriptions of electroweak defects, the structure of 
domain walls in non-Abelian theories, and the spectrum of 
magnetic monopoles in SU(5) Grand Unified theory. 
}

\vfill
\eject

\section*{Contents}

\begin{itemize}
\item[I.] Introduction
\begin{enumerate}
\item[A.] Friedman-Robertson-Walker cosmology
\item[B.] Cosmological phase transitions
\item[C.] Vacuum manifold
\end{enumerate}
\item[II.] Domain Walls
\begin{enumerate}
\item[A.] Topology: $\pi_0$
\item[B.] Example: $Z_2$ 
\item[C.] SU(5)
\item[D.] Formation
\item[E.] Evolution and constraints
\item[F.] Further complexities
\end{enumerate}
\item[III.] Strings
\begin{enumerate}
\item[A.] Topology: $\pi_1$
\item[B.] Example: U(1) local 
\item[C.] Semilocal and electroweak strings
\item[D.] Formation and Evolution
\item[E.] One scale model
\item[F.] Cosmological constraints and signatures
\begin{enumerate}
\item{} CMB and P(k)
\item{} Gravitational wave background
\item{} Gravitational lensing
\item{} Other signatures 
\end{enumerate}
\item[H.] Zero modes and superconducting strings
\end{enumerate}
\item[IV.] Magnetic Monopoles (in brief)
\begin{enumerate}
\item[A.] Topology: $\pi_2$
\item[B.] Examples: SU(2)
\item[C.]  Electroweak monopoles
\item[D.] SU(5) 
\item[D.] Cosmology (extremely brief)
\end{enumerate}
\begin{enumerate}
\item[A.] Further reading
\end{enumerate}
\end{itemize}
\vfill
\eject

\section{Introduction}
\label{introduction}

\subsection{FRW cosmology}
\label{FRWcosmology}

As indicated by observations, the universe is assumed to be 
homogeneous and isotropic on large scales. The line element is:
\begin{equation}
ds^2 = dt^2 - a^2(t) \biggl [ {{dr^2}\over {1-kr^2}}
               + r^2 ( d\theta^2 + \sin^2\theta d\phi^2 ) \biggr ]
\label{frwlineelement}
\end{equation}
where the function $a(t)$ is known as the scale factor, and
the parameter $k=-1,0,+1$ labels an open, flat or closed
universe. The Einstein equations give:
\begin{equation}
\biggl ( {{\dot a}\over a} \biggr ) ^2 + {k\over {a^2}}
= {{8\pi G}\over 3} \rho (t) 
\label{aeqn}
\end{equation}
where $\rho (t)$ is the energy density of the cosmological
fluid. Conservation of energy-momentum yields
\begin{equation}
\dot \rho + 3 H(\rho + p ) = 0
\label{dotrho}
\end{equation}
where $H \equiv {\dot a}/a$ is the Hubble parameter and
$p$ denotes the pressure of the cosmological fluid. The
properties of the cosmological fluid are given by its equation
of state:
\begin{equation}
p = p(\rho )
\label{eqnofstate}
\end{equation}
which must be specified to determine the expansion rate of
the universe. If the fluid is relativistic ({\it eg.} photons):
$p = \rho /3$; while if it is dust: $p=0$; and for vacuum
energy (cosmological constant): $p=-\rho$.

If the fluid is in thermal equilibrium, the evolution of the
temperature of the fluid can be derived by using the conservation
of entropy which gives
\begin{equation}
a(t) T = {\rm constant} \ .
\label{temperature}
\end{equation}
The temperature of the cosmic microwave background radiation
({\it i.e.} cosmological photons as opposed to those produced 
by stars and other astrophysical sources) at the present
epoch is measured to be $2.7^\circ$K.

It is sometimes convenient to work with the ``conformal
time'' $\tau$ instead of the ``cosmic time'' $t$. The
relation defining $\tau$ in terms of $t$ is:
\begin{equation}
dt = a(t) d\tau \ .
\label{conformaltime}
\end{equation}
In terms of the conformal time, the FRW line element is:
\begin{equation}
ds^2 = a^2(t(\tau )) \biggl [ d\tau^2 - {{dr^2}\over {1-kr^2}}
            + r^2 ( d\theta^2 + \sin^2\theta d\phi^2 ) \biggr ] \ .
\label{taufrwlineelement}
\end{equation}

\subsection{Cosmological phase transitions}
\label{cphasetransitions}

We consider a field theory of scalar, spinor and vector fields:
\begin{equation}
L = L_B+L_F
\label{fieldtheory}
\end{equation}
with the bosonic Lagrangian:
\begin{equation}
L_B = {1\over 2} D_\mu\Phi_i D^\mu\Phi_i - V(\Phi) 
    -{1\over 4} F_{\mu\nu}^aF^{\mu \nu a}
\label{bosonicL}
\end{equation}
where $\Phi_i$ are the components of the scalar fields.
The fermionic Lagrangian for a fermionic multiplet
$\Psi$ is:
\begin{equation}
L_F = i{\bar \Psi} \gamma^\mu D_\mu \Psi - 
           {\bar \Psi}\Gamma_i \Psi \Phi_i \ .
\label{fermionicL}
\end{equation}
In addition we have the following definitions:
\begin{equation}
D_\mu \equiv \partial_\mu -ie A_\mu^a T^a
\label{covariantD}
\end{equation}
where the $T^a$ are group generators;
\begin{equation}
F_{\mu\nu}^a \equiv \partial_\mu A_\nu^a - \partial_\nu A_\mu^a
                ++e f^{abc} A_\mu^b A_\nu^c
\label{fieldstrengthdefn}
\end{equation}
where, $A_\mu^a$ are the gauge fields.

If the expectation values of the scalar fields are denoted by
$\Phi_{0i}$, then
the mass matrices of the various fields are written as:
\begin{equation}
\mu_{ij}^2 = {{\partial^2 V}\over{\partial\Phi_i \partial\Phi_j}}
              \biggr | _{\Phi =\Phi_0} \ , \ \ \ {\rm scalar \ fields}
\label{mu2ij}
\end{equation}
\begin{equation}
m = \Gamma_i \Phi_{0i} \ , \ \ \ {\rm spinor \ fields}
\label{mspinor}
\end{equation}
where, the $\Gamma_i$ are the Yukawa coupling matrices, and
\begin{equation}
M^2_{ab} = e^2 (T_a T_b )_{ij} \Phi_{0i}\Phi_{0j} 
                      \ , \ \ \ {\rm vector \ fields}
\label{M2vector}
\end{equation}

Then the finite temperature, one-loop effective potential is:
\begin{equation}
V_{eff} (\Phi_0 , T) = V(\Phi_0 ) + {{{\cal M}^2} \over {24}}T^2
                     -{{\pi^2}\over {90}}{\cal N}T^4
\label{Veff}
\end{equation}
where
\begin{equation}
{\cal N} = {\cal N}_B + {7\over 8}{\cal N}_F
\label{calN}
\end{equation}
is the number of bosonic and fermionic spin states, and
\begin{equation}
{\cal M}^2 = {\rm Tr} {\mu^2} + 3{\rm Tr} {M^2} + 
            {1\over 2} {\rm Tr} (\gamma^0 m \gamma^0 m ) \ .
\label{calM2}
\end{equation}
Note that ${\cal M}^2$ depends on the expectation value 
$\Phi_0$ through the defining equations for the mass matrices
given above. Then, for example, ${\cal M}^2$ will contain a
term proportional to ${\rm Tr} (\Phi_0^2 )$.

For us the important feature of the effective potential is
that it can lead to cosmological phase transitions. If there
are scalar fields with negative mass squared terms in $V(\Phi )$,
the contributions from the ${\cal M}^2 T^2$ term in the effective
potential can make the effective mass squared positive for these
fields if the temperature is high enough. Therefore when the
universe is at a high temperature, the effective squared mass is
positive and the minimum of the potential is at $\Phi_0 =0$.
In the more recent cooler universe the effective mass squared
is negative and the minima of the effective potential will occur
at non-zero values of $\Phi_0$. That is, the scalar
fields will acquire vacuum expectation values.
This is the phenomenon of spontaneous symmetry breaking and,
if it occurs, will manifest itself as a cosmological phase transition.

\subsection{Vacuum manifold}
\label{vacuummanifold}

In general the field theoretic action under consideration will be 
invariant under some transformations of the fields. The set of
all such symmetry transformations form a group which we denote
by $G$. 

Next consider the situation where a scalar field acquires a
vacuum expectation value (VEV) $\phi_0$. Then all transformations
of $\phi_0$ by elements of $G$ will also be legitimate VEVs
of the scalar field. However, not all elements of $G$ will have
a non-trivial effect on $\phi_0$. There will be a subgroup of
$G$ which will have a trivial action on $\phi_0$. This is the
unbroken subgroup of $G$ and we denote it by $H$.

Now we wish to determine the set of possible VEVs of the scalar
field. This set is given by the elements of $G$ that have a
non-trivial action on $\phi_0$. Now consider an element of $g\in G$ 
that has a non-trivial action on $\phi_0$ and transforms it to
$\phi_1$. Then every element of the form $gh$ where $h\in H$,
also takes $\phi_0$ to $\phi_1$. Therefore the non-trivial 
transformations are given by the left cosets of $H$, and the
space of all non-trivial transformations is called a coset
space and denoted by $G/H$. Therefore the vacuum manifold of
the theory is $G/H$.

As we shall see, topological defects occur due to non-trivial 
topology of the vacuum manifold. 

Note that the topology does not care if the symmetries are local
or global. For most of these lectures, I will consider local
(gauged) symmetries.

\section{Domain walls}
\label{domainwalls}

\subsection{Topology: $\pi_0$}
\label{topologypi0}

Domain walls occur when the vacuum manifold has two or more
disconnected components. For example, $G$ can be a discrete
group like $Z_2$ which can be broken completely ($H={\bf 1}$). 
Then $G/H$ consists of just two points.

The topology of various manifolds has been studied by
mathematicians and they characterize the type of topology
by stating the homotopy groups of the manifold. The $n^{th}$
homotopy group of the manifold $G/H$ is denoted by 
$\pi_n (G/H)$ and denotes the group that is formed by
considering the mapping of $n-$dimensional spheres into the 
manifold $G/H$. Each element of the group $\pi_n$ is the
set of mappings that can be continuously deformed into
one another. 

In the case when the vacuum manifold has disconnected components,
$\pi_0 (G/H)$ is non-trivial since there are points
(zero dimensional spheres) that lie in different components
that cannot be continuously deformed into one another. Therefore
domain walls occur whenever $\pi_0 (G/H)$ is non-trivial.

\subsection{Example: $Z_2$}
\label{domainwallexample}

Consider the $Z_2$ Lagrangian in 1+1 dimensions labeled by $(t,z)$
\begin{equation}
L = {1\over 2}(\partial _\mu \phi )^2 - {\lambda\over 4} (\phi^2 -\eta^2 )^2
\label{z2model}
\end{equation}
where $\phi (t,z)$ is a real scalar field - also called the
order parameter. The Lagrangian
is invariant under $\phi \rightarrow -\phi$ and
hence possesses a $Z_2$ symmetry. For this
reason, the potential has two minima: $\phi = \pm \eta$,
and the ``vacuum manifold'' has two-fold degeneracy.

Consider the possibility that $\phi = +\eta$ at
$z= +\infty$ and $\phi = -\eta$ at $z=-\infty$. In
this case, the continuous function $\phi (z)$ has
to go from $-\eta$ to $+\eta$ as $z$ is taken from
$-\infty$ to $+\infty$ and so must necessarily
pass through $\phi =0$. But then there is energy in
this field configuration since the
potential is non-zero when $\phi =0$. Also, this configuration
cannot relax to either of the two vacuum configurations, say
$\phi (z) = + \eta$, since that involves changing the
field over an infinite volume from $-\eta$ to $+\eta$,
which would cost an infinite amount of energy.

Another way to see this is to notice the presence of a
conserved current:
$$
j^\mu = \epsilon^{\mu \nu} \partial_\nu \phi
$$
where $\mu, \nu =0,1$ and $\epsilon^{\mu \nu}$ is the
antisymmetric symbol in 2 dimensions. Clearly $j^\mu$
is conserved and so we have a conserved charge in the
model:
$$
Q = \int dz j^0 = \phi (+\infty ) - \phi (-\infty ) \ .
$$
For the vacuum $Q=0$ and for the configuration described
above $Q=1$. So the configuration cannot relax into the
vacuum - it is in a different topological sector.

To get the field configuration with the boundary
conditions $\phi (\pm \infty ) =\pm \eta$, one would have
to solve the field equation resulting from the Lagrangian
(\ref{z2model}). This would be a second order differential equation.
Instead, one can use the clever method first derived by
Bogomolnyi \cite{Bog76} and obtain a first order differential
equation. The method uses the energy functional:
\begin{eqnarray*}
E & = & \int dz
     [ {1\over 2}(\partial_t \phi)^2 + 
           {1\over 2}(\partial_z \phi)^2 + V(\phi ) ] \\
& = &  \int dz  [ {1\over 2}(\partial_t \phi)^2 +
                {1\over 2}(\partial_z \phi - \sqrt{2V(\phi )} ~ )^2 +
                  \sqrt{2V(\phi )}\partial_z \phi ]  \\
& = & \int dz [ {1\over 2}(\partial_t \phi)^2 +
{1\over 2} (\partial_z \phi - \sqrt{2V(\phi )} ~ )^2 ] +
\int^{\phi(+\infty )}_{\phi(-\infty )} d\phi ' \sqrt{2V(\phi ' )} \\
\end{eqnarray*}
Then, for fixed values of $\phi$ at $\pm \infty$, the energy is
minimized if
$$
\partial_t \phi =0
$$
and
$$
\partial_z \phi - \sqrt{2V(\phi )} = 0 \ .
$$
Furthermore, the minimum value of the energy is:
$$
E_{min} =  \int^{\phi(+\infty )}_{\phi(-\infty )}
d\phi ' \sqrt{2V(\phi ' )} \ .
$$
In our case,
$$
\sqrt{V(\phi )} = \sqrt{{\lambda\over 4}} (\eta^2 -\phi^2  )
$$
which can be inserted in the above equations to get the ``kink''
solution:
$$
\phi = \eta {\rm tanh} \biggl (\sqrt{\lambda\over 2} \eta z \biggr )
$$
for which the energy per unit area is:
\begin{equation}
\sigma_{kink} = {{2\sqrt{2}} \over 3} \sqrt{\lambda} \eta^3
              = {{2\sqrt{2}} \over 3} {{m^3}\over {\sqrt{\lambda}}}
\label{sigmakink}
\end{equation}
where $m = \sqrt{\lambda} \eta$ is the mass scale in the model
(see eq. (\ref{z2model})).
Note that the energy density is localized in the region where
$\phi$ is not in the vacuum, {\it i.e.} in a region of thickness
$\sim m^{-1}$ around $z=0$.

We can extend the model in eq. (\ref{z2model}) to 3+1 dimensions and
consider the case when $\phi$ only depends on $z$ but not on $x$ and
$y$. We can still obtain the kink solution for every value of $x$ and
$y$ and so the kink solution will describe a ``domain wall'' in the
$xy-$plane.

At the center of the kink, $\phi =0$, and hence the $Z_2$ 
symmetry is restored in the core of the kink. In this
sense, the kink is a ``relic'' of the symmetric phase
of the system. If kinks were present in the universe
today, their interiors would give us a glimpse of what
the universe was like prior to the phase transition.

\subsection{SU(5)}
\label{su5kink}

An example that is more relevant to cosmology is motivated
by Grand Unification. Here we will consider the SU(5) model:
\begin{equation}
L = {\rm Tr} ( D_\mu \Phi )^2 - 
      {1\over 2} {\rm Tr} (X_{\mu\nu} X^{\mu\nu}) -V(\Phi )
\label{su5bosoniclagrangian}
\end{equation}
where, in terms of components, $\Phi = \Phi^a T^a$ is an
SU(5) adjoint, the gauge field strengths are 
$X_{\mu\nu} = X_{\mu\nu}^a T^a$ and the SU(5) generators
$T^a$ are normalized such that ${\rm Tr} (T^aT^b) = \delta^{ab}/2$.
The definition of the covariant derivative is:
\begin{equation}
D_\mu \Phi = \partial_\mu \Phi -ie [X_\mu , \Phi]
\label{covariantderiv}
\end{equation}
and the potential is the most general quartic in $\Phi$:
\begin{equation}
V(\Phi ) = -m^2 {\rm Tr} (\Phi^2) + h [{\rm Tr} (\Phi^2)]^2 
   + \lambda {\rm Tr} (\Phi^4) + \gamma {\rm Tr} (\Phi^3) - V_0\ ,
\label{su5potential}
\end{equation}
where, $V_0$ is a constant that we will choose so as to set
the minimum value of the potential to zero.

The $SU(5)$ symmetry is broken to $[SU(3)\times SU(2)\times
U(1)]/Z_6$ if the Higgs acquires a VEV equal to
\begin{equation}
\Phi_0 = {\eta\over {2\sqrt{15}}} {\rm diag} (2,2,2,-3,-3)
\label{phi0}
\end{equation}
where
\begin{equation}
\eta = {m \over {\sqrt{\lambda '}}} \ , \ \ \ 
\lambda ' \equiv h+ {7\over {30}}\lambda \ .
\label{lambdaprime}
\end{equation}
For the potential to have its global minimum at $\Phi =\Phi_0$,
the parameters are constrained to satisfy:
\begin{equation}
\lambda \ge 0 \ , \ \ \ \lambda ' \ge 0 \ .
\label{constraint1}
\end{equation}
For the global minimum to have $V(\Phi_0 )=0$, in
eq. (\ref{su5potential}) we set
\begin{equation}
V_0 = - {{\lambda '}\over 4} \eta^4 \ .
\label{V0}
\end{equation}

The model in eq. (\ref{su5bosoniclagrangian}) does not 
have any topological
domain walls because there are no broken discrete symmetries. In
particular, the $Z_2$ symmetry under $\Phi \rightarrow -\Phi$ 
is absent due to the cubic term. However if $\gamma$ is small, 
there are walls connecting the two vacua related by 
$\Phi \rightarrow -\Phi$ that are almost topological. 
In our analysis we will set $\gamma =0$, in which case the symmetry 
of the model is $SU(5)\times Z_2$ and an expectation of $\Phi$ breaks 
the $Z_2$ symmetry leading to topological domain walls.

The kink solution is the Z$_2$ kink along the $\Phi_0$
direction (see eq. (\ref{phi0})). Therefore:
\begin{equation}
\Phi_k =  \tanh ( \sigma z ) \Phi_0
\label{phik}
\end{equation}
with $\sigma \equiv m/\sqrt{2}$ (see eq. (\ref{lambdaprime})),
and all the gauge fields vanish.
It is straightforward to check that $\Phi_k$ solves the
equations of motion with the boundary conditions
$\Phi (z=\pm\infty ) = \pm \Phi_0$.

The energy per unit area of the kink is (see eq. (\ref{sigmakink})):
\begin{equation}
M_k = \frac{2\sqrt{2}}{3} {{m^3}\over {\lambda '}} \ .
\label{kinkmass}
\end{equation}

The existence of a static solution to the equations of motion only
guarantees that it is an extremum of the energy but this extremum
may not be a minimum. To determine if the kink is a minimum energy
solution we need to examine its stability under arbitrary
perturbations. (As far as I know, a Bogomolnyi type analysis has
not been constructed for the SU(5) model.)

Here we will examine the stability of the kink under general perturbations.
So we write:
\begin{equation}
\Phi = \Phi_k + \Psi
\label{phiphikpsi}
\end{equation}
Since the kink solution is invariant under translations
and rotations in the $xy-$plane, it is easy to show that the
perturbations that might cause an instability arise from perturbations
of the scalar field and can only depend on $z$. Therefore we may set
the gauge fields to zero and take $\Psi =\Psi (t,z)$.
%\begin{equation}
%D_x \Phi D_x \Phi + D_y \Phi D_y \Phi + \sum_{i=1}^{3}(
%\frac{1}{2} X_{ix} X_{ix} + \frac{1}{2} X_{iy} X_{iy}) \, \ ,
%\end{equation}
%where $X_{ij}=\partial_i X_j - \partial_j X_i - i g [X_i,X_j]$ are
%the spatial components of the field strengths.
%This expression is non-negative and is minimized by setting the derivatives
%along the $x-$ and $y-$directions as well as the $x-$
%and $y-$components of the gauge fields to zero. We can also set the
%$z-$components of the gauge fields to zero because the remaining field
%strengths are vanishing identically.

The Z$_2$ kink is stable and hence we can restrict the
scalar perturbations to be orthogonal to $\Phi_k$. 
The perturbative stability of the kink has 
been studied by {\it Dvali et. al.} \cite{DvaLiuVac98},
and, {\it Pogosian and Vachaspati} \cite{PogVac00}.
Here we only consider the off-diagonal perturbation 
\begin{equation}
 \Psi  = \psi T \equiv \psi {\rm diag} (\tau^1 , 0,0,0) \ ,
\label{orthogonal}
\end{equation}
where $\tau^1$ is the first Pauli spin matrix.

Next we analyze the linearized Schrodinger equation for small excitations
$\psi=\psi_0 (z) exp(-i\omega t)$ in the background of the kink:
\begin{equation}
[- \partial^2_z - m^2+\phi_k^2(z)(h+\lambda r)] \psi_0 =
\omega_i^2 \psi_0 \, \ ,
\label{excitations}
\end{equation}
where $\phi_k \equiv \tanh (\sigma z)$ and $r=7/30$. 
The kink is unstable if there is a solution to
eq. (\ref{excitations}) with a negative $\omega^2$.
Substituting eq. (\ref{phik}) into eq. (\ref{excitations}) yields:
\begin{equation}
[-\partial^2_z+m^2(\tanh^2(\sigma z)-1)] \psi_0 =
\omega^2 \psi_0 \, \ .
\end{equation}
This equation has a bound state solution $\psi_0 \propto $ sech$(\sigma z)$
with the eigenvalue $\omega^2=-m^2/2$. Since this result is independent
of the parameters in the potential, we conclude that the kink in SU(5)
is always unstable.

So we still need to find the topological domain wall solution
in the model.

The domain wall solution is obtained if we choose the
gauge fields to vanish at infinity and the scalar field
to satisfy the boundary conditions:
\begin{eqnarray}
\Phi (z=-\infty )= \Phi^- \equiv
  {\eta \over {2\sqrt{15}}}  {\rm diag}(3,-2,-2,3,-2) \nonumber \\
         = \eta \sqrt{\frac{5}{12}} (\lambda_3+\tau_3)-
              \frac{\eta}{6}(Y-\sqrt{5}\lambda_8)
\label{phiminus}
\end{eqnarray}
and
\begin{eqnarray}
\Phi (z=+\infty ) = \Phi^+ \equiv
  {\eta \over {2\sqrt{15}}} {\rm diag}(2,-3,2,2,-3) \nonumber \\
         = \eta \sqrt{\frac{5}{12}} (\lambda_3+\tau_3)+
              \frac{\eta}{6}(Y-\sqrt{5}\lambda_8) \ .
\label{phiplus}
\end{eqnarray}
Here $\lambda_3$, $\lambda_8$, $\tau_3$ and $Y$ are the
diagonal generators of SU(5):
\begin{equation}
\lambda_3=\frac{1}{2} {\rm diag}(1,-1,0,0,0) \ ,
\label{lambda3}
\end{equation}
\begin{equation}
\lambda_8=\frac{1}{2\sqrt{3}} {\rm diag}(1,1,-2,0,0) \ ,
\label{lambda8}
\end{equation}
\begin{equation}
\tau_3=\frac{1}{2} {\rm diag}(0,0,0,1,-1) \ ,
\label{tau3}
\end{equation}
\begin{equation}
Y=\frac{1}{2\sqrt{15}} {\rm diag}(2,2,2,-3,-3) \ .
\label{ymatrix}
\end{equation}
Alternately, we could have chosen the boundary conditions to be
like those of the kink with $\Phi(z=+\infty )=-\Phi(z=-\infty )$.
However, then the solution for
the domain wall will not be diagonal at all $z$. We prefer to
use the above boundary conditions so that the solution is
diagonal throughout.

The domain wall solution can be written as
\begin{equation}
\Phi_{DW}(z)=a(z)\lambda_3+b(z)\lambda_8+c(z)\tau_3+d(z)Y \ .
\label{dwsolution}
\end{equation}
The functions
$a$, $b$, $c$, and $d$ must satisfy the static equations of motion:
\begin{eqnarray}
a''=[-m^2+(h+\frac{2\lambda}{5})d^2 &+& (h+\frac{\lambda}{2})(a^2+b^2)
+hc^2 ] a  \nonumber \\
&+&\frac{2\lambda abd}{\sqrt{5}}
\label{aequation}
\end{eqnarray}
\begin{eqnarray}
b''=[-m^2+(h+\frac{2\lambda}{5})d^2 &+& (h+\frac{\lambda}{2})(a^2+b^2)
+hc^2 ] b \nonumber \\
&+& \frac{\lambda d}{\sqrt{5}}(a^2-b^2)
\label{bequation}
\end{eqnarray}
\begin{eqnarray}
c''=[-m^2+(h+\frac{9\lambda}{10})d^2 +(h&+&\frac{\lambda}{2})c^2
                  \nonumber \\
&+& h(a^2+b^2)]c
\label{cequation}
\end{eqnarray}
\begin{eqnarray}
d''=[-m^2 &+& (h+\frac{7\lambda}{30})d^2+(h+\frac{2\lambda}{5})(a^2+b^2)
          \nonumber\\
        &+&(h+\frac{9\lambda}{10})c^2 ] d +
        \frac{\lambda b}{\sqrt{5}}(a^2-\frac{b^2}{3}) \ ,
\label{dequation}
\end{eqnarray}
where primes refer to derivatives with respect to $z$.
For reference, the kink solution (eq. (\ref{phik}))
corresponds to $a(z)=0=b(z)=c(z)$ and $d(z)= \eta \tanh(\sigma z)$.

The equations of motion for $b$ and $c$ and can be solved quite
easily:
\begin{equation}
b(z) = - \sqrt{5} d(z) \ , \ \ \ c(z)=a(z) \ .
\label{bcsoln}
\end{equation}
This is consistent with the boundary conditions in eqs. (\ref{phiminus})
and (\ref{phiplus}). In addition, we require
\begin{equation}
a(z=\pm \infty ) = + \eta \sqrt{5\over {12}} \ , \ \ \
d(z=\pm \infty ) = \pm {\eta \over 6} \ .
\label{adbcs}
\end{equation}
Then the remaining equations 
can be written in a cleaner form by rescaling:
\begin{equation}
A(z) = \sqrt{{12}\over 5} {a\over \eta} \ , \ \ \
D(z) = 6 {d\over \eta} \ , \ \ \
Z = mz \ .
\label{rescalings}
\end{equation}
This leads to 
\begin{equation}
A'' = \biggl [ -1 + {{(1-p)}\over 5} D^2 +
                     {{(4+p)}\over 5}A^2 \biggr ] A
\label{Aeqn}
\end{equation}
\begin{equation}
D'' = \biggl [ -1 + p D^2 + (1-p) A^2 \biggr ] D
\label{Deqn}
\end{equation}
where primes on $A$ and $D$ denote differentiation with respect
to $Z$, and
\begin{equation}
p = {1\over 6} \biggl [ 1+ {{5\lambda}\over {12\lambda '}}\biggr ] \ .
\label{pdefn}
\end{equation}
Note that $p\in [1/6,\infty )$ because of the constraints in
eq. (\ref{constraint1}).
The boundary conditions now are:
\begin{equation}
A(z=\pm \infty ) = +1 \ , \ \ \  D(z=\pm \infty ) = \pm 1 \ .
\label{newbcs}
\end{equation}

\begin{figure}
\begin{center}
\epsfxsize =  0.5 \hsize  \epsfbox{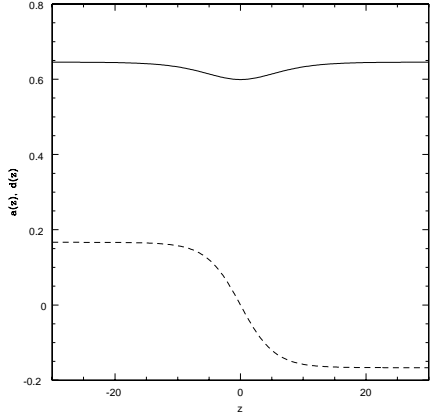}
\end{center}
\vskip 0.5 truecm
\caption{\label{danda}
Numerical solution for the domain wall in the case $\lambda =1$,
$h=-0.2$ ($p=2.25$). The solid line shows $a(z)$ and the dashed
line shows $d(z)$.
}
\end{figure}

This system of equations has been solved by numerical relaxation
and a sample solution is shown in Fig. \ref{danda}.
To find an approximate analytical solution, assume
that $|A''/A| << 1$ is small everywhere. This assumption
will be true for a certain range of the parameter $p$
which we can later determine.
Then the square bracket on the right-hand side of eq. (\ref{Aeqn})
is very small. This gives:
\begin{equation}
A \simeq \biggl [
          {{5\over {4+p}}} \biggl \{ 1 - {{(1-p)}\over 5} D^2
               \biggr \} \biggr ]^{1/2}
\label{Aapproxsoln}
\end{equation}
We insert this expression for $A$ in eq. (\ref{Deqn}) and
obtain the kink-type differential equation:
\begin{equation}
D'' = q[-1+D^2]D \ ,
\label{approxDeqn}
\end{equation}
where
\begin{equation}
q = {{6p-1}\over {p+4}} = {{6\lambda}\over {\lambda+60\lambda '}}
\label{qdefn}
\end{equation}
and the solution is:
\begin{equation}
D(Z) \simeq \tanh \biggl ( \sqrt{q\over 2} Z \biggr )
\label{Dapproxsoln}
\end{equation}
The parameter $q$ lies in the interval $[0,6]$. For $q=1$
({\it i.e.} $p=1$) it is easy to check that this analytical
solution is exact.

We can now check that our assumption $|A''/A| << 1$ is
self-consistent provided $p$ is not much larger than a few.

The energy density for the fields $A$ and $D$ can be found
from the Lagrangian in eq. (\ref{su5bosoniclagrangian}) together 
with the ansatz in eq. (\ref{dwsolution}), the solution for $b$ 
and $c$ in eq. (\ref{bcsoln}) and the rescalings in
eq. (\ref{rescalings}). The resulting expression for the
energy per unit area of the domain wall is:
\begin{equation}
M_{DW}= {{m^3}\over {12\lambda '}} \int dZ
         [ 5{A'}^2 + {D'}^2 + V(A,D) ]
\label{tension}
\end{equation}
where,
\begin{eqnarray}
V(A,D) = - 5A^2 &-& D^2 + {{(p+4)}\over 2} A^4 \nonumber \\
          &+& {p\over 2} D^4 + (1-p)A^2 D^2 + 3  \ .
\label{VAD}
\end{eqnarray}
The energy can be found numerically. However, here we will
find an approximate analytic result. We can insert the approximate
solution given above in eq. (\ref{tension}) but this leads to
an expression that is not transparent. Instead it is more useful
to consider another approximation for $A$ and $D$:
\begin{equation}
A \simeq 1 \ , \ \ \
D \simeq \tanh \biggl ( \sqrt{p\over 2} Z \biggr ) \ .
\label{approx2}
\end{equation}
(This approximation is exact for $p=1$.)
A straightforward evaluation then gives:
\begin{equation}
M_{DWapprox} = M_k {{\sqrt{p}} \over 6}
\label{sigmadw}
\end{equation}
where, $M_k$ is given in eq. (\ref{kinkmass}).

It can be shown for a range of parameters that this domain 
wall solution is perturbatively stable. Numerically we
find that it is lighter than the kink for all values of $p$.
(Eq. (\ref{sigmadw}) shows it to be lighter for $p<36$.)

The interior of the domain wall has less symmetry than the
exterior. So the domain wall does not contain a trapped
region of the early universe. Instead it is a region that
may be to our future where the $SU(3)$ symmetry is broken
down to $SU(2)\times U(1)$. In this sense, the domain wall
is a relic from the early universe but not a relic of the
unbroken symmetry of the theory.

\subsection{Formation}
\label{domainwallformation}

The properties of the network of domain walls at formation has 
been determined by numerical simulations. The idea behind the
simulations is that the
vacuum in any correlated region of space is determined
at random. Then, if there are only two degenerate vacuua
(call them + and -), there will be spatial regions
that will be in the + phase and others in the - phase.
The boundaries between these regions of different phases is 
the location of the domain walls.

By performing numerical simulations, the 
statistics shown in Table I was obtained \cite{VacVil84}.

\begin{table}[t]
\caption{
Size distribution of + clusters found by simulations on a
cubic lattice.
}\vspace{0.4cm}
\begin{center}
\begin{tabular*}{12.0cm}{|c@{\extracolsep{\fill}}ccccccc|}
\hline
{Cluster size} & {1} & {2} & {3} & {4} & {6} & {10} & {31082} \\[0.20cm]
%&&&&&&&&\\[-0.25cm]
\hline
%&&&&&&&&\\[-0.1cm]
{Number} & {462} & {84} & {14} & {13} & {1} & {1} & {1} \\[0.20cm]
\hline
\end{tabular*}
\end{center}
\end{table}

The data shows that there is essentially one giant connected + cluster.
By symmetry there will be one connected - cluster. In the infinite
volume limit, these clusters will also be infinite and their surface
areas will also be infinite. Therefore the topological domain wall 
formed at the phase transition will be infinite.

\subsection{Evolution}
\label{domainwallevolution}

Once the inifinite domain wall forms, it tries to straighten out since
it has tension -- in this way, the wall can minimize its energy. In the 
cosmological setting, the wall will be stretched out by the Hubble 
expansion. In addition, the wall will interact with the ambient matter 
and suffer friction. Sometimes different parts of the walls will collide
and interact. The interaction leads to the reconnection of colliding 
walls.

The details of the evolution of the walls are quite complicated.
However, the simplest cosmological scenario is easy to analyze
and leads to an interesting bound. The basic idea is to use
causality to get an upper bound on the energy density in domain
walls at any epoch. Causality tells us that the different domains
that are separated by walls cannot smooth out faster than the
speed of light. Therefore every causal horizon must contain at
least one domain wall at any epoch and so the energy density in
domain walls $\rho_{DW}$ obeys the following bound:
\begin{equation}
\rho_{DW} \ge {{\rm energy \ in \ one \ wall \ in \ horizon}\over 
             {\rm horizon\  volume}}
            \simeq {{\sigma t^2} \over {t^3}} = {{\sigma}\over t}
\label{rhodw}
\end{equation}
where $\sigma$ is the energy per unit area of the wall.
If we require that the energy density in walls be less than
the present critical density of the universe 
$\rho_{cr} = 3H^2/8\pi G$ with $H \simeq 70$ km/s/Mpc 
we get the constraint
\begin{equation}
\sigma < {{3H^2 t_0}\over {8\pi G}} \ ,
\label{sigmaconstraint}
\end{equation}
where $t_0 \sim 10^{17}$ secs is the present epoch.

If we use eq. (\ref{sigmakink}) to connect $\sigma$ to the symmetry 
breaking scale and take the coupling constant to be order one,
we find that the symmetry breaking scale is constrained to
be less than about a GeV. 
A stronger bound is obtained by realizing that the energy
density fluctuations in domain walls should not cause fluctuations
in the cosmic microwave background radiation (CMBR) greater than
1 part in $10^5$. This leads to the constraint that $\eta$ should 
be less than a few MeV.  Particle physics at such low energy scales
-- such as the standard model -- do not show any phase transitions
involving domain walls and so we do not expect to have domain
walls in the universe. Another way to view the constraint is
that if a particle physics model predicts domain walls above
the few MeV scale, cosmology rules out the model.

\subsection{Further complexities}
\label{complexities}

One can evade the constraints derived in the previous section
by introducing some new cosmological and particle physics
elements. 

For example, the constraint that domain walls 
have to be formed below the few MeV scale does not hold if
cosmological inflation follows the phase transition in which
heavy domain walls were produced. In such a scenario, the
domain walls are inflated away and our present horizon need
not contain any domain wall. 

In particle physics model building, one might have a situation
where a discrete symmetry is broken at some high temperature
and then restored at some lower temperature. In this case,
heavy domain walls would be formed at the first phase transition
and then would ``dissolve'' at the second phase transition.
Then there would be a period in the early universe where
domain walls would be present but they would not be around
today.

Another situation of some interest is when the discrete
vacuua are not exactly degenerate. In the SU(5) example
discussed above, if $\gamma$ (eq. (\ref{su5potential})) is 
very small, domain walls will be formed at the phase 
transition. Suppose the + phase has slightly less energy
density than the - phase. Then the domain walls will
experience a pressure on them that will drive them towards
the higher energy - phase. However, if $\gamma$ is small,
this pressure difference is also small and is negligible
for the early evolution of the wall system. Only after
the walls have straightened out to an extent that the
curvature forces are less than the pressure, will the
pressure start to drive the walls and eventually eliminate
them.

\section{Strings}
\label{strings}

\subsection{Topology: $\pi_1$}

If the vacuum manifold has one dimensional closed paths that cannot
be contracted, there are topological string solutions in the
field theory. The homotopy group $\pi_1 (G/H)$ is the group
formed by the equivalence classes of paths that can be deformed
into each other and the group operation joins two paths to
get another path. Each element of $\pi_1 (G/H)$ labels a
topologically distinct string solution.

An example of a field theory in which string solutions
exist is based on a $U(1)$ (global or local) group which
is broken down completely. The coset space $U(1)/{\bf 1}$
is a one dimensional sphere (circle). Closed paths that 
wrap around the circle cannot be contracted to a point
and this implies that the $U(1)$ model contains string
solutions. 

The $U(1)$ model is very important as it is relevant to the 
simplest superfluids and superconductors. However, in more
complicated systems the symmetry groups are larger and 
calculating the homotopy group can be more involved.
Fortunately there is a result that simplifies matters
immensely for most (though not all!) particle physics 
applications. This is the result that if $\pi_n (G)$ 
and $\pi_{n-1}(G)$ are both trivial then, 
\begin{equation}
\pi_n (G/H) = \pi_{n-1} (H) \ .
\label{homotopytheorem}
\end{equation}
With $n=1$ this gives
\begin{equation}
\pi_1 (G/H) = \pi_0 (H)
\label{nequal1homotopy}
\end{equation}
provided $\pi_1 (G) ={\bf 1}=\pi_0 (G)$. So if $G$ does
not contain any incontractable closed paths and does
not have disconnected pieces, then the topologically
distinct incontractable paths of $G/H$ are given by the 
disconnected pieces of $H$. So one can usually tell that 
there are strings in a particle physics model simply by 
checking if the symmetry breaking involves a broken 
$U(1)$ or if the unbroken group contains a discrete symmetry.
This result cannot be applied to the electroweak model since
there we have $G =[ SU(2)\times U(1) ]/Z_2$
and $\pi_1 (G) = {\bf Z} \ne {\bf 1}$.

\subsection{Example: U(1) local}
\label{stringexamples}

Consider the Lagrangian
$$
L = {1\over 2} |D_\mu \Phi |^2 - {1\over 4} F_{\mu \nu} F^{\mu \nu}
     - {\lambda \over 8}( \Phi^{*} \Phi - \eta^2 ) ^2
$$
where $\Phi$ is a complex field and
$$
D_\mu = \partial_\mu - ieA_\mu
$$
where, $A_\mu$ is an Abelian gauge potential.
This model has a U(1) gauge symmetry since it is invariant
under 
\begin{equation}
\Phi \rightarrow e^{i\theta}\Phi \ , \ \ \
A_\mu \rightarrow A_\mu + {1\over e} \partial_\mu \theta
\label{U(1)gaugetransform}
\end{equation}
Now since $\pi_1 (U(1)/{\bf 1}) = {\cal Z}$, the model has string
solutions labeled by an integer (the winding number).

The string solutions correspond to the non-trivial windings of
a circle at spatial infinity on to the vacuum manifold. Therefore
the boundary conditions that will yield the string solution will
have the form:
\begin{equation}
\Phi (r=\infty , \theta ) = \eta e^{in\theta} \ , \ \ \ 
                            n \in {\cal Z}-\{0\}
\label{Phiatinfty}
\end{equation}
The gauge fields at infinity must be such that the covariant
derivative vanishes there:
\begin{equation}
D_i \Phi = 0 \ , \ \ \ {\rm at} \ \ \ r=\infty \ .
\label{Aatinfty}
\end{equation}

Let us now construct the string solution using 
Bogomolnyi's method \cite{Bog76}.
The energy for static configurations in two spatial dimensions is 
\begin{equation}
E = \int d^2 x \biggl [ {1\over 2} |D_x\Phi |^2 + {1\over 2} |D_y\Phi |^2
                                  +{1\over 2} B_z ^2 + V(\Phi ) \biggr ] 
\label{U1energy}
\end{equation}
with
\begin{equation}
V(\Phi ) = {\lambda \over 8} ( |\Phi |^2 -\eta^2 )^2
\label{vphidefn}
\end{equation}
The trick is to write this as:
\begin{equation}
E = \int d^2 x \biggl [ {1\over 2} |D_x\Phi \pm iD_y \Phi|^2 + 
                        {1\over 2} (B_z - \sqrt{2V} )^2 \biggr ] +
                        {e\over 2}\eta^2  \int d^2 x B_z  
\label{bogoU1energy}
\end{equation}
in the special case that $\lambda =e^2$ (so called ``critical
coupling'') \footnote{Unlike in the case of the $Z_2$ domain wall, the
Bogomolnyi trick over here works only for the form of the potential
given in eq. (\ref{vphidefn}) and with critical coupling.}.
The Bogomolnyi equations are:
\begin{equation}
D_x \Phi + i D_y \Phi =0
\label{bogoeq1}
\end{equation}
\begin{equation}
B_z -\sqrt{2V} = 0
\label{bogoeq2}
\end{equation}
and the energy per unit length of the string is:
\begin{equation}
\mu = {e \over 2}\eta^2 \int d^2 x B_z = 
    {e \over 2}\eta^2 \oint d\theta A_\theta =
    \pi \eta^2 \ .
\label{bogoenergy}
\end{equation}
For non-Bogomolnyi strings, this expression will be multiplied
by a factor which depends on $\lambda/e^2$. Unless this parameter
is very large or small, numerical evaluations show that the numerical 
coefficient is of order unity.

For strings formed at the Grand Unified phase transition 
$\eta \sim 10^{16}$ GeV, we find $\mu \sim 10^{22}$ gms/cm which is
like the mass of a mountain range packed into a centimeter of string.
When calculating the gravitational effects of strings $\mu$ always
appears in the dimensionless combination $G\mu$. For GUT strings,
$G\mu \sim 10^{-6}$.

\subsection{Semilocal and electroweak strings}
\label{semilocalelectroweak}

The standard model symmetry breaking is $SU(2)_L \times U(1)_Y
\rightarrow U(1)_{em}$ and does not give topological strings.
However, it still contains embedded strings which can be 
perturbatively stable for a range of parameters which are
not realized in Nature. In these lectures I will only briefly
describe semilocal strings \cite{VacAch91}. 

Consider the generalization of the Abelian
Higgs model in which the complex scalar field is replaced by an
$SU(2)$ doublet $\Phi^T \ = \ ( \phi_1, \phi_2 )$.  The action is 
\begin{equation}
S = \int d^4 x \left [ | (\partial_\mu - iq Y_\mu )  \Phi |^2 \ - \
{1 \over 4} Y_{\mu \nu} Y^{\mu \nu}\ - \ {\lambda } \biggl (
\Phi^{\dag} \Phi - {\eta^2\over 2} \biggr ) ^2 \right ] \ ,
\label{semilocalaction}
\end{equation}
where  $Y_{\mu} $ is the $U(1)$ gauge potential and $Y_{\mu
\nu}= \partial_{\mu}Y_{\nu} - \partial_{\nu}Y_{\mu}$ its field
strength. 

This model has symmetry under 
$ G = [SU(2)_{global} \times U(1)_{local}] /Z_2$. Elements of the
$SU(2)_{global}$ act on the Higgs doublet, while elements of
$U(1)_{local}$ multiply the doublet by an overall phase and 
transform the gauge field in the usual manner. The action of the
center of $SU(2)_{global}$ on the Higgs doublet is identical to
a $U(1)_{global}$ phase rotation by $\pi$. This is the reason for
the $Z_2$ identification.
The model in eq. (\ref{semilocalaction}) is just the scalar sector 
of the Glashow-Salam-Weinberg (GSW) electroweak model with 
$\sin^2\theta_w = 1$. 

Once $\Phi$ acquires a vacuum expectation value, the symmetry breaks 
down to $H = U(1)_{global}$, as in the GSW model. The 
vacuum manifold is $S^3$. This may also be seen by minimizing
the potential explicitly. The vacuum manifold is described by
$\Phi^{\dag}\Phi = \eta^2 /2$ which is a three sphere. Hence
there are no incontractable paths on the vacuum manifold.
Yet, it is possible to perform a Bogomolnyi type analysis
to find that the configuration:
\begin{equation}
\Phi = f(\rho ) e^{i\theta} \pmatrix{0\cr 1} \ , \ \ \ 
Y_\theta = {{v(\rho )} \over {q\rho}}
\label{slsoln}
\end{equation}
is a solution in cylindrical coordinates $(\rho , \theta )$. 
The profile functions $f(\rho )$ and $v(\rho )$ can be
determined by solving the Bogomolnyi equations.

It is also possible to show that the solution
is perturbatively stable when $2\lambda < q^2$, neutrally
stable at critical coupling ($2\lambda =q^2$) and unstable
for $2\lambda > q^2$.

For the case when the SU(2) is gauged, the string solution
continues to exist and can be stable for 
$\sin ^2\theta_w > 0.9$ if the scalar mass is small. 
In the standard model, however,
$\sin ^2\theta_w =0.23$ and the string solution is certainly
unstable. Certain external conditions ({\it eg.} a magnetic
field) can stablize the string.

\subsection{Formation and evolution}
\label{formationevolution}

The formation of strings has been studied by numerical methods.
Here one throws down $U(1)$ phases on a lattice and checks
for topological winding around the plaquettes. If the winding
around a plaquette is non-trivial, it means that a string is
passing through it. Then the strings are connected and in this
way one obtains a network of closed and open strings.

The lattice simulations show several interesting properties
of the initial network of strings. First and foremost is the
result that most (over 80\%) of the string density is in strings 
that are infinite. Secondly, the string loops occur in a scale 
invariant distribution -- that is, the number density of loops 
having size between $R$ and $R+dR$ is proportional to $dR/R^4$, 
just as would be expected based on dimensional analysis. Thirdly,
the long strings have a Brownian shape and so the length $l$
of string in a distance $R$ is given by: $l \propto R^2$.

There has been much effort in extending the lattice simulations
to make them more realistic and closer to what might actually
happen at a phase transition. However, the essential results of 
the lattice simulations have held up very well. There has also
been a lot of theoretical and experimental effort devoted to 
determining the density of defects after the phase transition.
(In the simulations the density of defects is set by the lattice
spacing.) The statistical properties of the network described
above are expected to be insensitive to the initial string
density. The exact fraction of strings in infinite strings
might vary somewhat.

Once the string network has formed, it evolves under the
forces that we discussed in the domain wall case: (i) tension,
(ii) friction with ambient matter, (iii) Hubble expansion,
(iv) intercommuting (reconnection) when strings collide, and 
(iv) energy loss mechanisms such as gravitational wave emission. 
Of these factors, frictional forces are important for a short period
(for GUT strings) since the Hubble expansion dilutes the
ambient matter. The energy loss mechanisms are important
but to calculate their effect on the string dynamics is a
nasty back-reaction problem. The ``zeroth order'' factors
are the tension, Hubble expansion and intercommutings.

Ignoring friction, energy loss and intercommutings for now,
the strings move according to the Nambu-Goto action in a
background FRW spacetime:
\begin{equation}
S = -\mu \int d^2\sigma \sqrt{-g^{(2)}}
\label{NambuGotoaction}
\end{equation}
where $g^{(2)}$ is the determinant of the two dimensional
string world-sheet metric. If the string worldsheet is
written as: $x^\mu (\sigma^a)$ with $a=0,1$ where 
$(\sigma^0 ,\sigma^1)$ are the worldsheet coordinates, then
the worldsheet metric is:
\begin{equation}
g^{(2)}_{ab} = g_{\mu\nu}\partial_a x^\mu \partial_b x^\nu
\label{worldsheetmetric}
\end{equation} 
where, $g_{\mu\nu}$ is the four dimensional spacetime metric
in which the strings move.

The freedom of choosing coordinates on the worldsheet 
(reparametrization invariance) can be used to simplify
the equations of motion for the string derived from
eq. (\ref{NambuGotoaction}). We choose
\begin{equation}
\sigma^0 = \tau \ , \ \ \ \sigma^1 = \zeta
\label{parameterchoice}
\end{equation}
where $\zeta$ is a parameter along the string at any
given instant of the conformal time $\tau$ 
(see eq. (\ref{conformaltime})). Then the equation of motion is:
\begin{equation}
{\ddot {\bf x}} + 2H (1-{\dot {\bf x}}^2){\dot {\bf x}}
  = \epsilon^{-1} \biggl ( {{{\bf x}'}\over {\epsilon}} \biggr ) '
\label{eqofmotion}
\end{equation}
where overdots refer to derivatives with respect to 
$\tau$ and primes with respect to $\zeta$.

The equation of motion for a single string allows numerical
evolution of the network of strings but does not take intercommutings
into account. In practice, the network is numerically
evolved until two strings intersect and then the reconnection
is done by hand.

To study the evolution of cosmic strings by solving the equations
of motion is similar to studying a box of gas by solving Newton's
equations of motion for each particle in the box and taking care
of collisions by hand. For almost all cosmological purposes, only
the statistical properties of the string network are relevant.
So one should define ``statistical variables'' and find their
evolution. The one scale model for string evolution is precisely
such an attempt.

\subsection{One scale model}
\label{onescalemodel}

As the name suggests, the ``one scale model'' is based on the
assumption that there is only one length scale that characterizes 
the infinite strings in the network \cite{Kib85,Ben86}. 
(The short loops are considered
separately.) If we denote this scale by $L$ then the
distance between strings and the correlation length of a single
string (the distance out to which it is straight) are both given
by $L$. With time $L$ can change and hence we need to derive
an equation for $L(t)$.

The usual way of defining the length $L$ is by considering the
total energy $E$ in strings lying inside a volume $V$. Then
\begin{equation}
\rho \equiv {E \over V} \equiv {\mu\over {L^2}}
\label{defnofl}
\end{equation}
defines $L$. The energy changes
with time since the Hubble expansion dilutes and stretches the 
string network. Also, the velocity of a string redshifts due to 
the Hubble expansion. Long strings also lose energy by 
self-intersecting to form short loops. This gives
\begin{equation}
{\dot \rho} = -2 H (1+ v_{rms}^2 )\rho - c {\rho \over L}
\label{rhoequation}
\end{equation}
where 
\begin{equation}
v_{rms}^2 \equiv \langle v^2 \rangle 
              = {{\int d\zeta \epsilon {\dot {\bf x}}^2}
                           \over {\int d\zeta \epsilon}}
\label{vrms2}
\end{equation}
is the average rms velocity of strings
and $c$ is a parameter that accounts for the frequency of 
self-intersection (the so called ``chopping efficiency''). 
Equation (\ref{rhoequation}) can also be written as an
equation for $L$:
\begin{equation}
{\dot L} = HL (1+v_{rms}^2) + {c \over 2}
\label{Lequation}
\end{equation}
(Since the chopping term must vanish if the velocity of
the strings vanishes, one frequently writes 
$c={\tilde c} v_{rms}$.)

The network is said to scale if, at every epoch, the network
has the same statistical properties. Hence the scale $L$ should
be a fixed fraction of the horizon size $t$. In this case, the
energy density $\rho$ in long strings will fall off like
$1/t^2$, exactly as the total matter density does in a radiation
or matter dominated universe. In a universe with a cosmological 
constant, scaling will hold in a trivial sense since the rapid
expansion will dilute the strings to vanishing energy density.  
Scaling will not hold in the epochs during the period in which
the universe changes from being radiation dominated to matter 
dominated, or from matter dominated to cosmological constant
dominated. Both these transitions are relevant for examining
the cosmological signatures of strings ({\it eg.} the cosmic
microwave background anisotropies), making the problem quite
hard. 

To check for scaling we write
\begin{equation}
L = \gamma (t) t \ ,
\label{gammat}
\end{equation}
insert this relation into eq. (\ref{rhoequation}) to get
\begin{equation}
{{\dot \gamma}\over \gamma} = {1 \over t} \biggl [
 {c\over {2\gamma}} - \{ 1-Ht (1+ v_{rms}^2 ) \}
                                          \biggr ]
\label{gammadot}
\end{equation}
There is a fixed point solution (${\dot \gamma} = 0$)
at the point:
\begin{equation}
\gamma =
{c\over 2} {1 \over {1- (Ht) (1+v_{rms}^2 )}} \ .
\label{gammafixedpoint}
\end{equation}
Note that for a power law expansion $Ht$ does not depend
on time. The rms velocity could depend on time though we
can prove that $v_{rms}^2 =1/2$ in flat spacetime 
suggesting that it may be a reasonable assumption to take
it to be roughly constant even in the expanding universe.
However, we can derive an equation for $v_{rms}$ by 
differentiating eq. (\ref{vrms2}) with respect to 
time and then using the equation of motion. This gives
\begin{equation}
{\dot v}_{rms} = (1-v_{rms}^2) \biggl [ {k\over L}
                  - 2H v_{rms} \biggr ]
\label{vrmseq}
\end{equation}
where $k$ is a parameter (the ``momentum parameter'') 
related to the typical radius of curvature of the strings. 
The second term gives the redshifting of the velocity
due to Hubble expansion while the first describes the
change in the velocity due to the tension in the string.

Equation (\ref{Lequation}) and (\ref{vrmseq}) are the
one scale model equations for the string network. These
equations depend on the chopping efficiency $c$ 
(or $\tilde c$) and the
momentum parameter $k$ that need to be specified. The
practice has been to use the values found in numerical
simulations. Martins and Shellard \cite{MarShe00} give
\begin{equation}
{\tilde c} =0.23 \pm 0.04
\label{tildecvalue}
\end{equation}
and a velocity dependent value of $k$
\begin{equation}
k(v) = {{2\sqrt{2}}\over {\pi}} {{1-8v^6}\over {1+8v^6}}
\label{kvalue}
\end{equation}

\section{Cosmological constraints and signatures}
\label{signatures}

Numerical evidence suggests that the density of strings 
scales with the expansion of the universe and so we
have: $\rho \sim \mu /t^2$. Compared to the critical
density in a radiation- or matter-dominated universe
$\rho_c \sim 1/Gt^2$, the string density is
at least a factor $\sim G\mu$ smaller at all times.
So the strings never come to dominate the universe.

\subsection{CMB and P(k)}
\label{cmbpk}

The energy-momentum of the network of strings causes
perturbations in the metric which can introduce fluctuations
in the cosmic microwave background. The metric produced
by a straight string is conical with deficit angle
$8\pi G\mu$ and photons arriving to us 
from either side of a moving string have a temperature 
difference of $\sim 8\pi G\mu v$ where $v$ is the
velocity component of the string transverse to the
direction of propagation of the photons.
Therefore, on dimensional grounds, the CMBR anisotropy
produced by strings is:
\begin{equation}
{{\delta T}\over T} \sim 8\pi G\mu
\label{roughanistropy}
\end{equation}
which for GUT strings is about what is observed.

Several sophisticated calculations of the CMBR anisotropy
induced by strings have been done. I would like to draw
a distinction between some of the analyses that have
been done for global strings and others that have been
done for local strings. Here I will only discuss the 
specific analysis done by Pogosian and me of CMBR
anisotropies produced by local strings \cite{PogVac99}.

First of all we need to find the energy-momentum tensor
of the string network as it evolves over time. For this
we adopt a model first developed by Albrecht, Battye and
Robinson \cite{AlbBatRob99}.
The string network at any time is taken to be a gas of straight 
string segments of length $L$ and moving with velocity $v_{rms}$.
Then we evaluate the 
energy-momentum tensor of the network. For the evolution,
we follow the one-scale model described earlier. A tricky
part of the problem is to allow some of the strings to
decay. This is done by eliminating some of the segments
with time in such a way as to preserve the scaling form
of the string energy density.
Once the energy-momentum of the string network is obtained,
we feed it into the program CMBFAST to evaluate the 
anisotropy and also the power spectrum of density fluctuations. 

\begin{figure}
\begin{center}
\epsfxsize =  0.5 \hsize \epsfbox{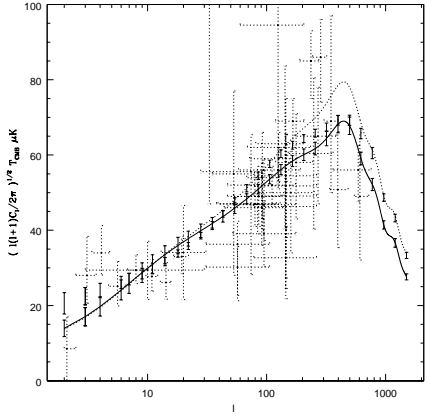}
\end{center}
\vskip 0.5 truecm
\caption{\label{clvsl}
The spectrum of microwave background fluctuations
for the network of wiggly strings with $L=0.1$. The 
solid and dotted lines correspond to two values of
the network parameters. The observations with error
bars are also shown.
}
\end{figure}

Some of the results are shown in Fig. \ref{clvsl}.
(I am assuming that you have already been introduced to the 
theory behind the microwave background measurements.)
Clearly there is only one Doppler peak and it occurs at
$l\sim 400$ and this does not agree with observations.

\begin{figure}
\begin{center}
\epsfxsize =  0.5 \hsize \epsfbox{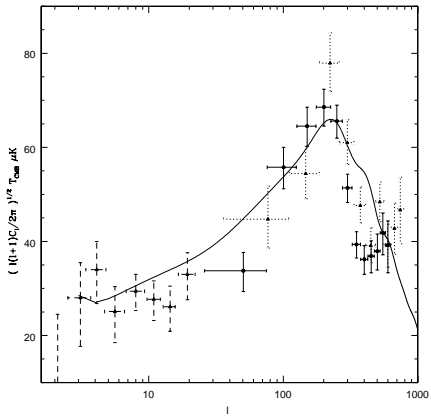}
\end{center}
\vskip 0.5 truecm
\caption{
\label{clclosed}
As in Fig. \ref{clvsl} but for a closed universe
with $\Omega =1.3$. See the article by 
{\it Pogosian} \protect \cite{Pog00} 
for fuller account of the parameters used and the
underlying assumptions.
}
\end{figure}

It is known that larger values of $\Omega$ bring down the
position of the Doppler peak. So Pogosian has recently 
investigated local strings in a closed model \cite{Pog00}. 
The result that 
the peak position only scales as $\Omega^{-1.58}$ would seem 
to imply that we would need to go to $\Omega \sim 3$
in order to bring down the peak position by a factor of
two or so. However, this result does not apply to perturbation
sources such as cosmic strings where the anisotropy is affected
by the gravitational potentials produced after last scattering.
Simulations by Pogosian \cite{Pog00} show that the peak position 
shifts to $l\sim 200$ for $\Omega \simeq 1.3$ 
(see Fig. \ref{clclosed}). 
Furthermore, the closest fit to the data occurs for 
$G\mu = 1-2\times 10^{-6}$. While this matches the GUT scale,
the poor fit to the data means that GUT strings cannot be
solely responsible for producing the required density
inhomogeneities.

\subsection{Gravitational wave background}
\label{gwb}

When local strings oscillate at very low frequencies, they 
can only radiate very low energy particles. Frequencies
set by cosmic scales are $\sim 1/t$ and the only known particles
that can be radiated are massless - photons and gravitons. 
(Neutrinos are assumed to have a non-zero mass.)
Generally the (local) strings are made of neutral fields and
hence do not radiate significantly in photons \footnote{If
the strings are superconducting they may carry electrical
charges and currents, and then they would lose energy by
emitting photons.}. Then the
primary decay channel for strings is into gravitons \footnote{
Global string loops decay primarily by emitting Goldstone bosons.}.

To estimate the power emitted in gravitational radiation by 
a string loop of length $L$, it is most convenient to use the 
quadrupole formula. The only dimensional parameters available to us
are $G$, $\mu$ and $L$. However, we know that the power loss into
gravitational waves is proportional to one factor of $G$ and two 
factors of $\mu$ since the power is the square of the quadrupole 
moment. Now since the power has dimensions of mass squared, we must 
have
\begin{equation}
P \sim G\mu^2
\label{Pestimate}
\end{equation}
and the length of the loop drops out from the estimate.
In the cosmic scenario we are interested in the average
gravitational energy emitted per unit time by all strings.
So we write:
\begin{equation}
{\bar P} = \Gamma G\mu^2
\label{barP}
\end{equation}
where $\Gamma$ is a numerical factor. By examining the power
emitted by many different loops, one finds $\Gamma \sim 100$
\cite{VacVil85}.

To find the present gravitational wave background amplitude 
and spectrum one needs to sum over all strings in the network,
include appropriate redshift factors etc.. The final
result is \cite{CalAll92}:
\begin{equation}
\Omega_g (\omega ) = {{18\pi^2 (\beta -1)^2 \nu G\mu} \over
                      {(3-\beta )\sin[(2-\beta )\pi]}} 
                    \biggl ( {{4\pi}\over {\Gamma\mu\omega t_0}}
                    \biggr )^{\beta-1}
\label{Omegag}
\end{equation}
where $\nu$ is a parameter that sets the amplitude for the number 
density of loops and $4/3 \le \beta << 2$ is a parameter that
characterizes the typical spectrum of gravitational radiation
emitted by loops, $t_0$ is the present epoch. The peak of the 
spectrum is at 
\begin{equation}
\omega_{peak} \sim {{4\pi}\over {\Gamma G\mu t_0}} \ .
\label{gwpeak}
\end{equation}

The strongest constraints on the amplitude of the gravitational 
wave background arise from the timing of the millisecond pulsar.
In 1993 the constraint was:
\begin{equation}
\Omega_g < 4\times 10^{-7} h^{-2}
\label{Omegagconstraint}
\end{equation}
The constraint gets more stringent the longer the millisecond
pulsar is observed without encountering noise in its timing.
Today the constraint would be slightly stronger than in 1993.
In terms of $G\mu$, this gives
\begin{equation}
G\mu < 4\times 10^{-5}
\label{Gmuconstraint}
\end{equation}

The emission of gravitational radiation from strings also
contributes to the energy density prior to nucleosynthesis.
If the gravitational radiation contribution to the content
of the universe exceeds about 5\% , it will interfere
unacceptably with BBN. This gives another constraint on 
the parameter $G\mu$ that is slightly better, though
comparable to, the one in eq. (\ref{Gmuconstraint}).

\subsection{Gravitational lensing}
\label{lensing}

Spacetime distortions due to cosmic strings would lead
to gravitational lensing of background sources. The
angular separation of images can be deduced from the
value of the conical deficit angle to be:
\begin{equation}
\delta \phi = 8\pi G\mu {d \over {d+l}}
\label{lensangle}
\end{equation}
where $d$ is the distance between the source and
the string and $l$ is the distance between string
and observer. This leads to an image separation of 
about 5 arc sec from GUT strings and is quite large 
by astronomical standards.

\begin{figure}
\begin{center}
\epsfxsize =  \hsize  \epsfbox{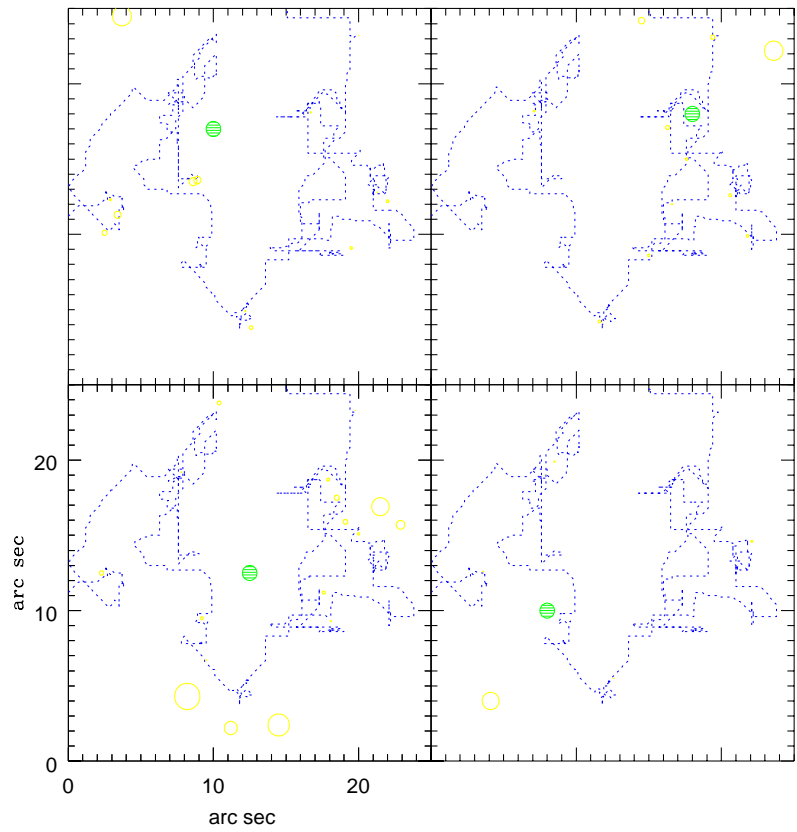}
\end{center}
\vskip 0.5 truecm
\caption{\label{stringlens}
The projection of a long wiggly string on the sky is 
shown together with a variety of point sources (hatched
circles). The images of the point sources are shown
as unfilled circles.
}
\end{figure}

Initially it was thought that a string would produce
a line of lensed images of sources and this would be
a unique signature. However, since the strings are
wiggly, this conclusion is no longer obvious. Instead
one needs to examine the lensing of sources due to
a more realistic (wiggly) string. This was done 
by {\it de Laix et. al.} \cite{deLKraVac97} and the results 
are shown in Fig. \ref{stringlens}.

\subsection{Other signatures}
\label{othersignatures}

Cosmic strings and other topological defects have been
considered as sources for ultra high energy cosmic rays
and gamma ray bursts. 

\section{Zero modes and superconducting strings}
\label{zeromodes}

The interaction of fermions with strings can often
lead to ``zero modes'' {\it i.e.} zero energy solutions
to the Dirac equation in the string background. If these
fermions carry electric charge, electric fields along
the string (equivalently, string motion through a 
magnetic field) can produce electric currents on the
string which will persist even when the electric
field is turned off. Hence such strings are said to
be ``superconducting''.

The current on a superconducting string depends on 
the initial current on it at formation and also on
the flux of magnetic field that it has traversed.
The motion of the current carrying string leads 
to electromagnetic radiation that can dissipate the
string energy. At the same time, the electromagnetic
radiation can cause spectral distortions in the
cosmic microwave background. This places constraints
on the density of superconducting strings in the
early universe and also the amount of current
on them if they exist.

\section{Magnetic monopoles}
\label{magneticmonpoles}

\subsection{Topology: $\pi_2$}
\label{topologypi2}

Magnetic monopoles are formed when the vacuum manifold
has incontractable two spheres. The relevant homotopy
group is $\pi_2 (G/H)$. As discussed in the string
case, if $\pi_2 (G)={\bf 1}=\pi_1(G)$, then 
\begin{equation}
\pi_2 (G/H)= \pi_1 (H) \ .
\label{pi2pi1}
\end{equation}
Hence we get magnetic monopoles whenever the {\it unbroken}
group contains incontractable paths. We know that today
the unbroken group contains the U(1) electromagnetic
gauge symmetry which contains incontractable paths.
Hence, any GUT predicts magnetic monopoles.

\subsection{Example: SU(2)}
\label{examples}

The 't Hooft-Polyakov monopole is based on an SU(2) model:
\begin{equation}
L = -{1\over 4} X^a_{\mu\nu}X^{a\mu\nu} +
     {1\over 2} (D_\mu\Phi^a )^2 - V(\Phi ) \ ,
\label{su2model}
\end{equation}
where $\Phi$ is an SU(5) adjoint scalar field,
$X^a_{\mu\nu}$ ($a=1,...,24$) are the gauge
field strengths and the covariant derivative is defined by:
\begin{equation}
D_\mu \Phi^a = \partial_\mu \Phi^a - i e [X_\mu ,\Phi]^a
\label{covariantderivative}
\end{equation}
and the group generators are normalized by
${\rm Tr}(T_a T_b)=\delta_{ab}/2$.
Once $\Phi$ acquires a VEV, the symmetry
is broken down to U(1), consisting of SU(2) rotations that
leave $\Phi$ invariant.

It is possible to give a Bogomolnyi type derivation of the
monopole solution in the case that the potential vanishes 
{\it i.e.} $V(\Phi ) =0$. The explicit
solution in spherical coordinates is \cite{PraSom75}:
\begin{equation}
\Phi = \frac{1}{er^2}(\frac{Cr}{\tanh(Cr)}-1)  x^a T_a 
\end{equation}
\begin{equation}
W^a_i = \epsilon^a_{~ij}
\frac{x^j}{er^2}\left ( 1-\frac{Cr}{\sinh(Cr)} \right ) \ . 
(a,i,j=1,2,3) \ ,
\end{equation}
This $V=0$ monopole is referred to as the Bogomolnyi-
Prasad-Sommerfield (BPS) monopole.
For $V \ne 0$, the profile functions can be found
numerically.

The mass of the BPS monopole also follows from the
Bogomolnyi analysis and is:
\begin{equation}
M = {{4\pi }\over {e^2}} m_V 
\label{bpsmass}
\end{equation}
where $m_V$ is the mass of the massive vector bosons
after symmetry breaking. In terms of the vacuum expectation
value $\eta$ of $\Phi$, $m_V = e\eta$.

\subsection{Electroweak monopoles}
\label{ewmonopoles}

The standard electroweak model does not contain the suitable
topology to have magnetic monopole solutions. This is because 
$\pi_1 (G) \ne {\bf 1}$ and the incontractable loops in the 
unbroken U(1) are also incontractable in $G$. Yet, as originally 
discovered by Nambu \cite{Nam77}, the model does contain magnetic 
monopoles that are confined by strings. (This is closely analagous 
to the picture of quarks being confined by QCD strings.)

The standard model has a doublet $\Phi$, SU(2) gauge
fields $W_\mu^a$ and the hypercharge gauge field $Y_\mu$.
The SU(2) and U(1) coupling constants are denoted by $g$ and
$g'$ and the vacuum expectation value of $\Phi$ is denoted
by $\eta$. The weak mixing angle $\theta_w$ is defined via
$\tan \theta_w = g'/g$.
The asymptotic configuration of the electroweak monopole can be 
written in spherical coordinates (with $r\rightarrow \infty$)
as:
\begin{equation}
\Phi = {{\eta} \over {\sqrt{2}}}
     \pmatrix{ \cos(\theta /2) \cr \sin(\theta /2) e^{i\varphi }} \
\label{ewmonophi}
\end{equation}
\begin{equation}
g W_\mu ^a = - \epsilon^{abc} n^b \partial_\mu n^c + i \cos^2 \theta_w
   n^a ({\Phi}^{\dag} ~ \partial_\mu {\Phi} -
        \partial_\mu {\Phi}^{\dag} ~{\Phi} )
\label{ewmonow}
\end{equation}
\begin{equation}
g' Y_\mu = - i \sin^2 \theta_w
 ({\Phi}^{\dag} ~ \partial_\mu {\Phi} -
   \partial_\mu {\Phi}^{\dag} ~{\Phi} ) \
\label{ewmonoy}
\end{equation}
where
\begin{equation}
n^a (x) \equiv  
  -{{\Phi^{\dag}(x) \tau^a \Phi (x)} \over {\Phi^{\dag}(x) \Phi}(x)}\ .
\label{nadef}
\end{equation}

Note that the configuration has a singularity along
the negative $z$ axis -- the lower component of $\Phi$
is not single-valued here. This is the location of a real
string. As described in Sec. \ref{semilocalelectroweak}, 
this string is the semilocal string when $\theta_w = \pi /2$
and the electroweak Z-string for general $\theta_w$.

The electroweak monopole is not a static {\it solution} of the
equations of motion but only a field configuration. One
way to see this is that to note that the string is pulling
on the monopole and this unbalanced force will tend to
accelerate the monopole and to shorten the string. Eventually
the monopole will annihilate the antimonopole at the other
end of the string. However this fact may not diminish the 
importance of the monopole -- the important aspect to consider 
is the lifetime of the electroweak monopole and antimonopole
and to evaluate if they can lead to some experimental signatures 
\cite{Nam77}. At the moment, it seems unlikely that the electroweak 
monopole can survive long enough to be seen in accelerator 
experiments though this issue needs further exploration.

\subsection{SU(5) monopoles}
\label{su5monopoles}

The Grand Unified symmetry breaking 
$SU(5) \rightarrow [SU(3)\times SU(2)\times U(1)]/Z_6$
leads to topological magnetic monopoles. 
The magnetic monopoles in this model have been
constructed \cite{DokTom80}. Here I will
not go into the details of the construction. Instead
I will only point out an interesting feature of the spectrum
of stable magnetic monopoles.

The winding one monopole in the model has SU(3),
SU(2) and U(1) charges. Then one can construct higher
winding monopoles by assembling together unit winding
monopoles. In some cases, the assembly will be stable
while in other cases the higher winding monopole will
be unstable to decaying into lower winding monopoles.
It turns out that, for a broad range of parameters,
the stable monopoles are the winding $\pm$1, $\pm$2, 
$\pm$3, $\pm$4, and $\pm$6 monopoles. The magnetic
charges on these can be calculated. As shown in
Table II an interesting feature is that the 
known quarks and leptons have precisely the same spectrum
of charges in the electric sector! This observation
has led to the possibility that it might be possible
to understand the fundamental particles as magnetic
monopoles \cite{Vac96}. 

\begin{table}[t]
\caption{
The quantum numbers ($n_8$, $n_3$ and $n_1$) on stable $SU(5)$
monopoles are shown and these correspond to the $SU(3)$, $SU(2)$
and $U(1)$ charges on the corresponding standard model fermions
shown in the right-most column.
}\vspace{0.4cm}
\begin{center}
\begin{tabular*}{8.0cm}{|c@{\extracolsep{\fill}}cccc|}
\hline
&&&&\\[-0.42cm]
                   {$n_{~}$}
                 & {$n_8 /3$}
                 & {$n_3 /2$}
                 & {$n_1 /6$}
                 & {$$} \\
%\tableline
\hline
&&&&\\[-0.42cm]
+1&1/3&1/2&1/6 &$(u,d)_L$    \\
-2&1/3&0  &-1/3&$d_R$        \\
-3&0  &1/2&-1/2&$(\nu ,e)_L$ \\
+4&1/3&0  &2/3 &$u_R$        \\
-6&0  &0  &-1  &$e_R$        \\
%\tableline
\hline
\end{tabular*}
\end{center}
\end{table}
\vspace{0.15cm}

\subsection{Cosmology of monopoles}
\label{cosmology}

Once magnetic monopoles are produced at a phase transition
in the early universe, their energy density will redshift
like pressureless matter $\rho_m \propto 1/a^3$. The radiation 
redshifts as $\rho_\gamma \propto 1/a^4$. Hence magnetic
monopoles will start dominating the universe at some epoch.
The exact epoch of domination will depend on the details of
monopole formation and their evolution. However, even if we
assume one monopole per causal horizon at formation, 
GUT monopoles would have started dominating well before 
big bang nucleosynthesis, which is clearly in conflict
with observations \cite{ZelKhl78,Pre79}.

Another bound on the monopole density arises from the 
observation that galaxies have magnetic fields that would
accelerate any magnetic monopoles that are present. In this
way the monopoles would dissipate the magnetic field. This
``Parker bound'' leads to a constraint that is stronger
than the cosmological constraint for lighter monopoles 
\cite{Par70}.

The present constraints on monopoles imply that magnetic
monopoles must be more massive than about $10^{10}$ GeV if 
they are to be present in cosmology. For certain types of
monopoles ({\it eg.} those that catalyse proton decay),
the constraints from stellar physics can be much stronger.

The constraints on magnetic monopoles are clearly in conflict
with the result that the GUT phase transition must have
produced magnetic monopoles. Hence either the GUT philosophy
is incorrect, or the standard FRW cosmology is incorrect.
The popular solution to the monopole over-abundance problem 
is that the standard cosmology should be modified to include 
an inflationary phase of the universe. Then the monopole producing 
GUT phase transition occurs during the inflationary stage and any 
monopoles that were produced are rapidly diluted to insignificant 
densities. Other possible solutions are to (i) modify particle physics
such that the monopoles get connected by strings and then annihilate,
\cite{LanPix80}
(ii) never have an epoch where the Grand Unified symmetry is restored
\cite{DvaMelSen95},
and (iii) produce domain walls that sweep away the monopoles
\cite{DvaLiuVac98}.

\section{Further reading}
\label{furtherreading}

The topological defects literature is vast and the citations 
have not done justice to all the work that has been done.
Fortunately a number of excellent review articles and books
have been written that can be a guide to the original literature.
Here are some of these articles:

\begin{itemize}

\item{} 
``Solitons and Particles'', eds. C. Rebbi
and G. Soliani (World Scientific, Singapore, 1984). This is
an invaluable resource for learning the subject and for reading 
most of the classic papers on the field theoretic aspects.

\item{}
``Cosmic Strings and Other Topological Defects'',
A. Vilenkin and E.P.S. Shellard, Cambridge University
Press (1994). This is a comprehensive description of
the field.

\item{}
M.B. Hindmarsh and T.W.B. Kibble,
Rept. Prog. Phys. {\bf 58}, 477 (1995). An excellent discussion
of the many different facets of cosmic strings. 

\item{}
P. Goddard and  D. I. Olive, Rep. Prog. Phys.
{\bf 41} 1357-1437 (1978). A classic on magnetic monopoles.

\item{}
S. Coleman, 
``The magnetic monopoles, fifty years later'',
in: The Unity of Fundamental Interactions, ed A. Zichichi (Plenum,
London, 1983). Another classic.

\item{}
J. Preskill, ``Vortices and monopoles'', in:
Architecture of Fundamental Interactions at Short Distance, 
eds P.  Ramond and R. Stora (Elsevier, 1987). And yet another.

\item{}
``Semilocal and Electroweak Strings'',
A. Ach\'ucarro and T. Vachaspati, Phys. Rep. {\bf 327}, 347 (2000).

\item{}
``Aspects of $^3$He and the Standard Electroweak Model'',
G.E. Volovik and T. Vachaspati, {\it Int. J. Mod. Phys.} {\bf B10}
471-521 (1996). A review of topological defects in He-3 with a
particle physics connection.

\end{itemize}

\section*{Acknowledgements}

This work was supported by the Department of Energy, USA.


\section*{References}

\begin{thebibliography}{99}

\bibitem{Bog76} E. B. Bogomolny, Sov. J. Nucl. Phys. {\bf 24}
449 (1976); reprinted in ``Solitons and Particles'', eds. C. Rebbi
and G. Soliani (World Scientific, Singapore, 1984).

\bibitem{DvaLiuVac98} 
G. Dvali, H. Liu and T. Vachaspati, 
Phys. Rev. Lett. {\bf 80}, 2281 (1998).

\bibitem{PogVac00}  L. Pogosian and T. Vachaspati,
Phys. Rev. {\bf D62}, 123506 (2000).

\bibitem{VacVil84}
T. Vachaspati and A. Vilenkin,
Phys. Rev. {\bf D30}, 2036 (1984).

\bibitem{VacAch91} T. Vachaspati and
A. Ach\'ucarro, {\it Phys. Rev.} {\bf D44}, 3067 (1991).

\bibitem{Kib85}
T.W.B. Kibble,
Nucl. Phys. {\bf B261}, 750 (1985).

\bibitem{Ben86}
D.P. Bennett,
Phys. Rev. {\bf D33}, 872 (1986); Erratum: Phys. Rev. {\bf D34},
3932 (1986).

\bibitem{MarShe00} C. J. A. P. Martins and E. P. S. Shellard,
hep-ph/0003298 (2000).

\bibitem{PogVac99} 
L. Pogosian and T. Vachaspati,
Phys. Rev. {\bf D60}, 083504 (1999).

\bibitem{AlbBatRob99}
A. Albrecht, R. Battye and J. Robinson,
Phys. Rev. {\bf D59}, 023508 (1999).

\bibitem{Pog00}
L. Pogosian, astro-ph/0009307 (2000).

\bibitem{VacVil85}  
T. Vachaspati and A. Vilenkin, 
Phys. Rev. {\bf D31}, 3052 (1985).

\bibitem{CalAll92}
R.R. Caldwell and B. Allen,
Phys. Rev. {\bf D45}, 3447 (1992).

\bibitem{deLKraVac97} 
A.A. de Laix, L.M. Krauss and T. Vachaspati, 
Phys. Rev. Lett. {\bf 79}, 1968 (1997).

\bibitem{PraSom75}
M.K. Prasad and C.H. Sommerfield,
Phys. Rev. Lett. {\bf 35}, 760 (1975).

\bibitem{Nam77}
Y. Nambu, Nucl. Phys. {\bf B130}, 505 (1977).

\bibitem{DokTom80}
C.P. Dokos and T.N. Tomaras,
Phys. Rev. {\bf D21}, 2940 (1980). 

\bibitem{Vac96}
T. Vachaspati, 
Phys. Rev. Lett. {\bf 76}, 188 (1996).

\bibitem{ZelKhl78}
Ya.B. Zeldovich and M.Yu. Khlopov,
Phys. Lett. {\bf B79}, 239 (1978).

\bibitem{Pre79}
J. Preskill,
Phys. Rev. Lett. {\bf 43}, 1365 (1979).

\bibitem{Par70}
E.N. Parker, Ap. J. {\bf 160}, 383 (1970).

\bibitem{LanPix80}
P. Langacker and S.-Y. Pi,
Phys. Rev. Lett. {\bf 45}, 1 (1980).

\bibitem{DvaMelSen95}
G. Dvali, A, Melfo and G. Senjanovic, Phys. Rev. Lett.
{\bf 75}, 4559 (1995)











%\bibitem{twbk} T. W. B. Kibble, {\it J. Phys.} {\bf A9}, 1387 (1976).


%\bibitem{guth} A. H. Guth, {\it Phys. Rev.} {\bf D23}, 347 (1981).

%\bibitem{preskill} J. Preskill, {\it Phys. Rev. Lett.} {\bf 43},
%1365 (1979).

%\bibitem{khuang} K. Huang, ``Statistical Mechanics'', (John Wiley and
%Sons, New York, 1987).

%\bibitem{ll} L. D. Landau and E. M. Lifshitz, 
%``Statistical Physics, Part I'', revised and enlarged 
%by E. M. Lifshitz and L. P. Pitaevskii (Pergamon Press, Oxford, 1980).

%\bibitem{peskinbook} M. E. Peskin and D. V. Schroeder, ``Introduction
%to Quantum Field Theory'' (Addison-Wesley, Reading, Mass., 1995).

%\bibitem{rr} R. J. Rivers, ``Path Integral Methods in Quantum
%Field Theory'', (Cambridge University Press, 1987).

%\bibitem{scew} S. Coleman and E. J. Weinberg, {\it Phys. Rev.}
%{\bf D7}, 1887 (1973). 

%\bibitem{linde} A. Linde, {\it Rep. Prog. Phys.} {\bf 42}, 389 (1979). 

%\bibitem{sw1} S. Weinberg, {\it Phys. Rev.} {\bf D9}, 3357 (1974).

%\bibitem{th} G. 't Hooft, {\it Nucl. Phys.} {\bf B79}, 276 (1974). 

%\bibitem{ap} A. M. Polyakov, {\it JETP Lett.} {\bf 20}, 194 (1974).

%\bibitem{avps} A. Vilenkin and E. P. S. Shellard, ``Cosmic Strings
%and Other Topological Defects'' (Cambridge University Press, 1994).

%\bibitem{herstein} I. N. Herstein, ``Topics in Algebra''
%(Xerox College Publishing, Lexington, Massachusetts, 1975).

%\bibitem{tvaa} T. Vachaspati and A. Ach\'ucarro,
%{\it Phys. Rev.} {\bf D44}, 3067 (1991).

%\bibitem{mhsemi} M. B. Hindmarsh, {\it Phys. Rev. Lett.} {\bf 68},
%1263 (1992).

%\bibitem{jpsemi} J. Preskill, {\it Phys. Rev.} {\bf D46}, 4218 (1992).

%\bibitem{rudazam} Rudaz and A. M. Srivastava,
%{\it Mod. Phys. Lett.} {\bf A8}, 1443 (1993).

%\bibitem{tvav} T. Vachaspati and A. Vilenkin, {\it Phys. Rev.} 
%{\bf {D30}}, 2036 (1984).

%\bibitem{scherrerfrieman} R. J. Scherrer and J. A. Frieman,
%{\it Phys. Rev.} {\bf D33}, 3556 (1986).

%\bibitem{lptv} L. Pogosian and T. Vachaspati, hep-ph/9709317, 
%CWRU-P14-97 (1997).

%\bibitem{leeseprokopec} R. A. Leese and T. Prokopec, {\it Phys. Lett.}
%{\bf B260}, 27 (1991).

%\bibitem{tvbias} T. Vachaspati, {\it Phys. Rev.} {\bf D44}, 3723 (1991).

%\bibitem{mhstrobl} M. Hindmarsh and K. Strobl, {\it Nucl.Phys.} {\bf B437}, 
%471 (1995).

%\bibitem{scherrervilenkin} R. J. Scherrer and A. Vilenkin,
%{\it Phys. Rev.} {\bf D56}, 647 (1997).

%\bibitem{twbkav} T. W. B. Kibble and A. Vilenkin, Phys. Rev. {\bf D52},
%679 (1995).

%\bibitem{bubcoll}
%A. Melfo and L. Perivolaropoulos, {\it Phys. Rev.} {\bf D52}, 992 (1995);
%P. Saffin and E. Copeland, {\it Phys. Rev.} {\bf D54}, 6088 (1996);
%A. Ferreira and A. Melfo, Phys. Rev. {\bf D53}, 6852 (1996);
%S. Digal, S. Sengupta and A. M. Srivastava, Phys. Rev. {\bf D56}, 
%2035 (1997);
%J. Ahonen and K. Enqvist, hep-ph/9704334 (1997). For related
%discussions, also see
%A. Yates and T. W. B. Kibble, {\it Phys.Lett.} {\bf B364}, 149 (1995); 
%J. Robinson and A. Yates, {\it Phys. Rev.} {\bf D54}, 5211 (1996);
%Scherrer and Vilenkin \cite{scherrervilenkin}.

%\bibitem{jbtktvav} J. Borrill, T. W. B. Kibble, T. Vachaspati
%and A. Vilenkin, {\it Phys. Rev.} {\bf D52}, 1934 (1995).
%
%\bibitem{jb} J. Borrill, {\it Phys. Rev. Lett.} {\bf 76}, 3255 (1996).

%\bibitem{nalbmh} N. D. Antunes, L. M. A. Bettencourt and M. Hindmarsh,
%hep-ph/9708215 (1997).
%
%\bibitem{aakklptv} A. Ach\'ucarro, K. Kuijken, L. Perivolaropoulos and 
%T. Vachaspati, {\it Nucl. Phys.} {\bf {B388}}, 45 (1992).
%
%\bibitem{aaetal} A. Ach\'ucarro, J. Borrill and A. R. Liddle,
%hep-ph/9702368 (1997).
%
%\bibitem{yokoyama} M. Nagasawa and J. Yokoyama, {\it Phys. Rev. Lett.}
%{\bf 77}, 2166 (1996).
%

\end{thebibliography}
\end{document}